\documentclass[printer,floatfix]{aa} 
\usepackage{graphicx}

\begin{document}
\def\al{&\!\!\!\!}
\def\x{{\bf x}}
\def\f{\frac}

\title{Distant star clusters of the Milky Way in MOND }

\author{ Hossein Haghi$^{1,2}$,
         \inst{} \fnmsep
         \thanks{\email{haghi@iasbs.ac.ir}}
          Holger Baumgardt$^{2,3}$,  \thanks{\email{h.baumgardt@uq.edu.au}}
        \inst{}
          \and
         Pavel Kroupa$^2$   \thanks{\email{pavel@astro.uni-bonn.de}}
        \inst{}          }
\offprints{H. Haghi}

 \institute{ $^{1}$Institute for Advanced Studies in Basic Sciences (IASBS), P.O. Box 11365-9161, Zanjan, Iran\\
$^{2}$Argelander Institute for Astronomy (AIfA), Auf dem H\"ugel 71,
D-53121 Bonn, Germany\\
$^{3}$School of Mathematics and Physics, University of Queensland, Brisbane, QLD 4072, Australia\\}

\date{Accepted}

\abstract{ We determine the mean velocity dispersion of six Galactic
outer halo globular clusters, AM 1, Eridanus, Pal 3, Pal 4, Pal 15,
and Arp 2 in the weak acceleration regime to test classical vs.
modified Newtonian dynamics (MOND). Owing to the nonlinearity of
MOND's Poisson equation, beyond tidal effects, the internal dynamics
of clusters is affected by the external field in which they are
immersed. For the studied clusters, particle accelerations are much
lower than the critical acceleration $a_0$ of MOND, but the motion
of stars is neither dominated by internal accelerations ($a_i \gg
a_e$) nor external accelerations ($a_e \gg a_i$). We use the N-body
code N-MODY in our analysis, which is a particle-mesh-based code
with a numerical MOND potential solver developed by Ciotti,
Londrillo, and Nipoti (2006) to derive the line-of-sight velocity
dispersion by adding the external field effect. We show that
Newtonian dynamics predicts a low-velocity dispersion for each
cluster, while in modified Newtonian dynamics the velocity
dispersion is much higher. We calculate the minimum number of
measured stars necessary to distinguish between Newtonian gravity
and MOND with the Kolmogorov-Smirnov test. We also show that for
most clusters it is necessary to measure the velocities of between
30 to 80 stars to distinguish between both cases. Therefore the
observational measurement of the line-of-sight velocity dispersion
of these clusters will provide a test for MOND.  }
\authorrunning{Haghi et al.}
\titlerunning{Globular clusters in MOND}
\maketitle

\section{Introduction}

Observable matter in galaxies and in clusters of galaxies cannot
produce sufficient gravity to explain their dynamics. Cold dark
matter (CDM) scenarios or alternative theories of gravitation are
therefore invoked to resolve the problem. Nowadays, the CDM
hypothesis is the dominant paradigm. This hypothetical matter does
not interact with electromagnetic radiation and only shows its
presence through its gravitational interaction. Even though the dark
matter hypothesis has successfully explained the internal dynamics
of galaxy clusters, gravitational lensing, and the standard model of
cosmology within the framework of general relativity (GR)
(\cite{spe03}), much experimental effort has failed to yield a
detection of dark matter particles. Moreover the results of
high-resolution simulations of structure formation do not reproduce
some observations on galactic scales, such as the central structures
of rotation curves, the prevalence of low bulge-to-disc ratios, and
the numbers and spatial distribution of the subhalos
(\cite{kly99,moore99,metz08, kroupa10}). Even the ability of the
dark matter theory to account for the Tully-Fisher and Freeman
relations is controversial (\cite{bos00,gov10}). These shortcomings
have not led to the rejection of the theory only because on galactic
scales baryons are at least non-negligible contributors to the mass
density, consequently simulations that include the complex physics
of star formation are essential for reliable predictions. Currently
such simulations are still at an experimental stage and are usually
substituted by ``semi-analytic'' arguments that have weak
theoretical underpinnings.

One of the alternative theories to CDM is the so-called modified
Newtonian dynamics (MOND) theory, which was originally proposed by
Milgrom (1983) to explain the flat rotation curves of spiral
galaxies at large distances by a modification of Newton's second law
of acceleration below a characteristic scale of
$a_{0}\simeq1.2\times10^{-10} $ms$^{-2}=3.6$ pc/(Myr)$^2$ without
invoking dark matter (\cite{bek84}). In MOND, Newton's second law is
modified to $\mu(a/a_{0}){\bf a} = {\bf a}_{N} + \nabla\times{\bf
H}$, where $\rho$ is the mass-density distribution, ${\bf a}_N$ is
the Newtonian acceleration vector, ${\bf a}$ is the MONDian
acceleration vector, $a=|\textbf{a}|$ is the absolute value of
MONDian acceleration, $\mu$ is an interpolating function for the
transition from the Newtonian to the MONDian regime, which runs
smoothly from $\mu(x)=x$ at $x\ll1$ to $\mu(x)=1$ at $x\gg1$
(\cite{bek84}). Different interpolating functions have been
suggested, such as the simple function, $\mu(x)=x/(1+x)$
(\cite{fam05}) and the
standard interpolation function, $\mu(x)=x/\sqrt{1+x^2}$
(\cite{mil83}). Because the simple function fits galactic rotation
curves better than the standard function (\cite{gent10}), in this
paper we use the simple function. The value of the curl field ${\bf
H}$ depends on the boundary conditions and the spatial mass
distribution and vanishes only for some special symmetries
(\cite{bek84}). The non-linearity of the MOND field equation leads
to difficulties for standard N-body codes and makes the use of the
usual Newtonian N-body simulation codes impossible in the MOND
regime.

It has been shown that on galactic scales MOND can explain many
phenomena better than CDM (\cite{beg89, begm91, san02, san07, hag06,
mal09, gen07, mil94, bra00, mil95, wu08, za06, tir07,has10}). MOND
has been generalized to a general-relativistic version
(\cite{bek04,san05,zlo07,mil09}), making it possible to test its
predictions for gravitational lensing. Dynamics of galaxies in
clusters (\cite{san02}) and the merging of galaxy clusters, where
the baryonic mass is clearly separated from the gravitational mass
(\cite{Clo06}) cannot be completely explained by MOND without
invoking some kind of hot dark matter, perhaps in the form of
massive (active or sterile, 2 to 11 eV) neutrinos
(\cite{ang06,ang10}).

In order to decide whether MOND is a comprehensive theory to explain
the dynamics of the universe, it is desirable to study MOND for
objects in which no dark matter is supposed to exist and where the
characteristic acceleration of the stars is less than the MOND
critical acceleration parameter $a_0$. Globular clusters (GCs) are a
perfect candidate since they are the largest virialized structure
that do not contain dark matter (\cite{moore96}).

In the distant halo of our Milky Way there exist several low-mass
GCs where both internal and external accelerations of stars are
significantly below the critical acceleration parameter $a_0$ of
MOND. Because GCs are assumed to be dark-matter-free, if MOND is
true, the motions of stars must deviate from the standard Newtonian
dynamics. It has been proposed by Baumgardt (2005) that some of
these distant Galactic GCs are perfect tools to test gravitational
theories in the regime of very weak accelerations. For MOND, the
internal velocity dispersion among the stars in these clusters would
be significantly higher than in Newtonian dynamics.

The mean velocity dispersion of stellar systems for the two extreme
cases of internal ($a_i\gg a_e$) or external ($a_e\gg a_i$) field
domination have been derived analytically by Milgrom
(\cite{mil86,mil94}), assuming that the systems are everywhere in
the deep-MOND regime ($a_e, a_i \ll a_0$). Many systems that can be
used to test MOND are not completely either internally or externally
dominated. Globular clusters or dwarf galaxies of the Milky Way for
example have internal and external accelerations that are of the
same order (\cite{bau05}), consequently one has to determine the
velocity dispersion numerically for intermediate cases. Sollima and
Nipoti (2009) constructed self-consistent, spherical models for
stellar systems in MOND, neglecting the external field effect and
presented a dynamical model for six galactic globular clusters. The
presence of the external field effect breaks the spherical symmetry
and validity of their model. Recently Haghi et al. (2009, hereafter
HBK09) investigated the dynamics of star clusters by numerically
modeling them in MOND, assuming circular orbits. They performed
N-body simulations and presented analytical formulae for the
velocity dispersion of stellar systems in the intermediate MOND
regime, which are useful for a comparison with observational data of
several GCs and dSph galaxies (for details on the numerical
calculations see HBK09). In a follow-up paper, Jordi et al. (2009)
determined the velocity dispersion (using 17 stars) and
mass-function slope of Pal 14 and showed that MOND can hardly
explain the low-velocity dispersion of this system. However, Gentile
et al. (2010) showed that with the currently available data, the
Kolmogorov-Smirnov (KS) test is still unable to exclude MOND with a
sufficiently high confidence level. Moreover, the low density of Pal
14 suggests that binary stars may be an important issue for
interpreting its measured velocity dispersion (\cite{Kuepper10}),
and the true velocity dispersion of Pal 14 could be much lower than
the value reported by Jordi et al. (2009), thereby possibly posing
an even larger challenge for MOND, but also for Newtonian gravity
and for any understanding of the dynamics of this object as being in
equilibrium.

In this paper we calculate the prediction of MOND and Newtonian
dynamics on the velocity dispersion of six other distant clusters of
the MW (Table 1). In order to see the pure MONDian effects, we
concentrate on systems in which the tidal radius is much larger than
the gravitational radius \footnote{ The gravitational radius is a
measure of the size of the system and is related to the mass and
potential energy as given in Eq. 2-132 in Binney and Tremaine
(1987). In many stellar systems the gravitational radius can be
approximated by the three-dimensional half-mass radius $r_h$ as
$r_g=1.25r_h$ if the assumption of virial equilibrium is valid. }
and therefore tidal effects are unimportant. In other words, this
paper provides the basis for further observational efforts. The
measurements of a low- (Newtonian) velocity dispersion would mean
that MOND in its present form is in severe trouble and that globular
clusters do not possess dark matter. In contrast, a high-velocity
dispersion would either favor MOND or could be a hint to the
existence of dark matter in globular clusters (\cite{bau09}).

The paper is organized as follows: In Section 2 we give a brief
review of the external field effect (EFE) in MOND. The simulation
setup is explained in Section 3. The numerical results for six
clusters are discussed in Section 4. We present our conclusions in
Section 5.

\section{External field effect in MOND}

In classical Newtonian dynamics, a uniform external field does not
affect the internal dynamics of a stellar system. In other words, in
the frame of the system relative motions of objects are the same as
in an isolated system.

In MOND, the situation is entirely different. Owing to the
non-linearity of Poisson's equation, the strong equivalence
principle (SEP) is violated (\cite{bek84}), and consequently the
internal properties and the morphology of a stellar system are
affected by both the internal and external field. This so-called
external field effect (EFE) significantly affects non-isolated
systems and can provide a strict test for MOND.

The EFE is indeed a phenomenological requirement of MOND and was
postulated by Milgrom (1983) to explain the dynamical properties of
open clusters in the MW that do not show MONDian effects. The EFE
allows high-velocity stars to escape from the potential of the Milky
Way (\cite{fam07,wu07}) and implies that rotation curves of spiral
galaxies should fall where the internal acceleration becomes equal
to the external acceleration (\cite{gen07,wu08}).

For a star cluster with a density distribution $\rho_{c}$, which is
embedded in a host galaxy with a density distribution $\rho_{ext}$,
the acceleration of stars in the cluster satisfies the modified
Poisson equation
\begin{equation}
 \nabla.[\mu(\frac{\nabla\Phi}{a_{0}})\nabla\Phi] = 4\pi G (\rho_c+\rho_{ext}), \label{mon1}
\end{equation}
where $\nabla\Phi$ is the MONDian potential generated by the total
matter density. For star clusters or dwarf galaxies far out in the
halo of the Milky Way the local density of the Milky Way is
negligible ( i.e. $\rho_{ext} << \rho_c$ ). One way to solve Eq. 1
is then to assume that the total acceleration is the sum of the
internal $a_i$ and the external $a_e$ acceleration, which both
satisfy the modified Poisson equation as
(\cite{bek84,wu07,wu08,ang08})

\begin{equation}
 \nabla.[\mu(\frac{|\bf{a_e}+\bf{a_i}|}{a_{0}})({\bf a_e}+{\bf a_i})] = 4\pi G \rho_c. \label{mon2}
\end{equation}

In Eq. 2, the direction of the external acceleration is important
for the dynamics of stars in the clusters. For example, for two
stars the acceleration would be different if $a_e$ is parallel or
anti-parallel to their internal acceleration.

Several attempts have been made considering the EFE. Wu et al.
(2007) simulated an isolated object with a static MONDian potential
solver, but changed the boundary condition on the outermost grid
point to be nonzero. Famaey et al. (2007) estimated the escape
velocity of galaxies in MOND and assumed
$\mu(|\bf{a_e}+\bf{a_i}|/a_{0}){\bf a_i}={\bf a_N}$. As a first
order test they replaced $|\bf{a_e}+\bf{a_i}|$ with $(a_e+a_i)$, or
$\sqrt{a_i^2+a_e^2}$. This is an approximation, since they neglected
the direction of $\bf{a_e}$ (i.e. a possible angular difference
between $\bf{a_i}$ and $\bf{a_e}$).

In order to study the evolution of the systems we need to calculate
the acceleration at each step by adding the constant external
gravitational field to the internal acceleration inside the
$\mu$-function, considering the different angle between external and
internal acceleration for all stars throughout the evolution.

\section{Simulation setup }

We performed a large set of N-body simulations of star clusters with
the N-MODY code, which has been developed by Ciotti et al. (2006).
N-MODY is a parallel, three-dimensional particle-mesh code for the
time-integration of collision-less N-body systems (\cite{lon09}).
The code numerically solves the non-linear MOND field equations,
which can be used to perform numerical experiments in either MONDian
or Newtonian dynamics. The potential solver of N-MODY is based on a
grid in spherical coordinates and is best suited for modeling
isolated systems. N-MODY uses the leap-frog method to advance the
particles. The code and the potential solver have been presented and
tested by Ciotti et al. (2006) and Nipoti et al. (2007).

In the present study we used a spherical grid ($r,\theta,\varphi$)
made of $N_r\times N_\theta \times N_\varphi=128\times128\times128$
grid cells for the integration. The total number of particles is
$N_p = 10^5$. The details of the scaling of the numerical MOND
models and code units are discussed in Nipoti, Londrillo \& Ciotti
(2007).

Our treatment of the external field effect is based on the idea that
the star clusters we study are much smaller than they are distant
from the Milky Way, so the external field is nearly constant over
the cluster area. We carry out the simulations in a rotating
reference frame centered on the cluster. We also assume that the
clusters are on a circular orbit, so the external field of the Milky
Way is also constant with time in the clusters frame.


%
%
%
%
We assume that tidal forces arising from a gradient of the external
field and the Coriolis acceleration caused by the rotating reference
frame can be neglected. This is justified as long as the size of the
cluster is much smaller than the tidal radius. We then solve Eq. 2
numerically with N-MODY, using a constant acceleration $a_e$ as the
boundary condition.

\begin{table*}
\begin{center}
\begin{tabular}{cccccccc}
Cluster name   &$R_h$[pc]&$R_{G}$[kpc]& $a_e$$[a_0]$   &$M_c$$[10^3$M$_{\odot}]$ &$\sigma_M$[kms$^{-1}$]&$\sigma_N$[kms$^{-1}$]& $N_{min}$\\
 \hline
      AM 1                     &24        &123.2         &0.086    &12.6 & 1.50              & 0.56  & 25            \\
      Eridanus                 &14.2      &95.2          &0.113    &18.6 & 1.80              & 0.90  & 40             \\
      Pal 3                    &24        &95.9          &0.112    &31.6& 1.97               & 0.87  & 30          \\
      Pal 4                    &23.2      &111.8         &0.096    &42.6 & 2.30              & 1.10  & 35          \\
      Pal 15                   &21.2      &37.9          &0.283    &26.3 & 1.42              & 0.88  & 80          \\
      Arp 2                    &21.5      &21.4          &0.504    &21.8 & 1.07              & 0.80  & 150          \\
\hline
\end{tabular}
\end{center}
\caption{Globular clusters modeled in this paper. The half-mass
radius and galactocentric distances, $R_G$, are taken from Harris
(1996). The values of the external acceleration are calculated from
$a_{e} = \sqrt{GMa_0}/R_G$  with $M=1.2 \times 10^{11}$M$_{\odot}$
for all galactocentric distances. Cluster masses, $M_c$, were
calculated from the absolute V-band luminosities by assuming a
stellar mass-to-light ratio of $M/L_V=2$, which is close to the
measured average mass-to-light ratio of galactic globular clusters
(Mieske et al. 2008). $\sigma_M$ and $\sigma_N$ are the
corresponding MONDian and Newtonian values of the velocity
dispersion, respectively. The last column is the minimum number of
stars necessary to obtain $P\leq 0.05$.  \label{table} }
\end{table*}


In order to include the EFE for non-isolated systems, we applied
several changes to the N-MODY code. The changes were encoded in the
source file \emph{mond-lib.f90}, which contains all relevant
routines implementing the MOND potential solver. The Mondian
potentials and accelerations are assigned to stars in subroutines:
"\emph{gmond}" and "\emph{mond-acc}", and we placed the vector of
$a_e$ in the interpolating function for all stars, i.e., $a_e$ is
added to $a_i$ via vector summation and the code solves Eq. 2 by
iteration. At each step of potential solving, we added the constant
external field for all grid points according to Eq. 2. This is a
first approximation that allows us to focus on the effects of a
constant external field, which in turn allows us to deal with the
intermediate MOND regime. We would like to stress that our numerical
solution agrees very well with analytic ones for the extreme cases
$a_i\ll a_e\ll a_0$, $a_e\ll a_i\ll a_0$ and $a_e,a_i\gg a_0$
(HBK09).

Our method is a first attempt to take into account the constant EFE
in N-MODY, in order to avoid solving the modified Poisson equation
for the large area including both galaxy and cluster, which is
impossible with the current version of N-MODY. This can only be
studied with a realistic, high-resolution MOND simulation that
includes the density distribution causing the external field.

In this work we start from Newtonian equilibrium Plummer models and
produce MONDian equilibrium initial systems. In order to have a
MONDian equilibrium initial system, we increased the initial
velocity of the particles. This method is useful when the external
field is important. The details of the method are described in
HBK09.

\begin{figure*}
\centering
\includegraphics[width=185mm]{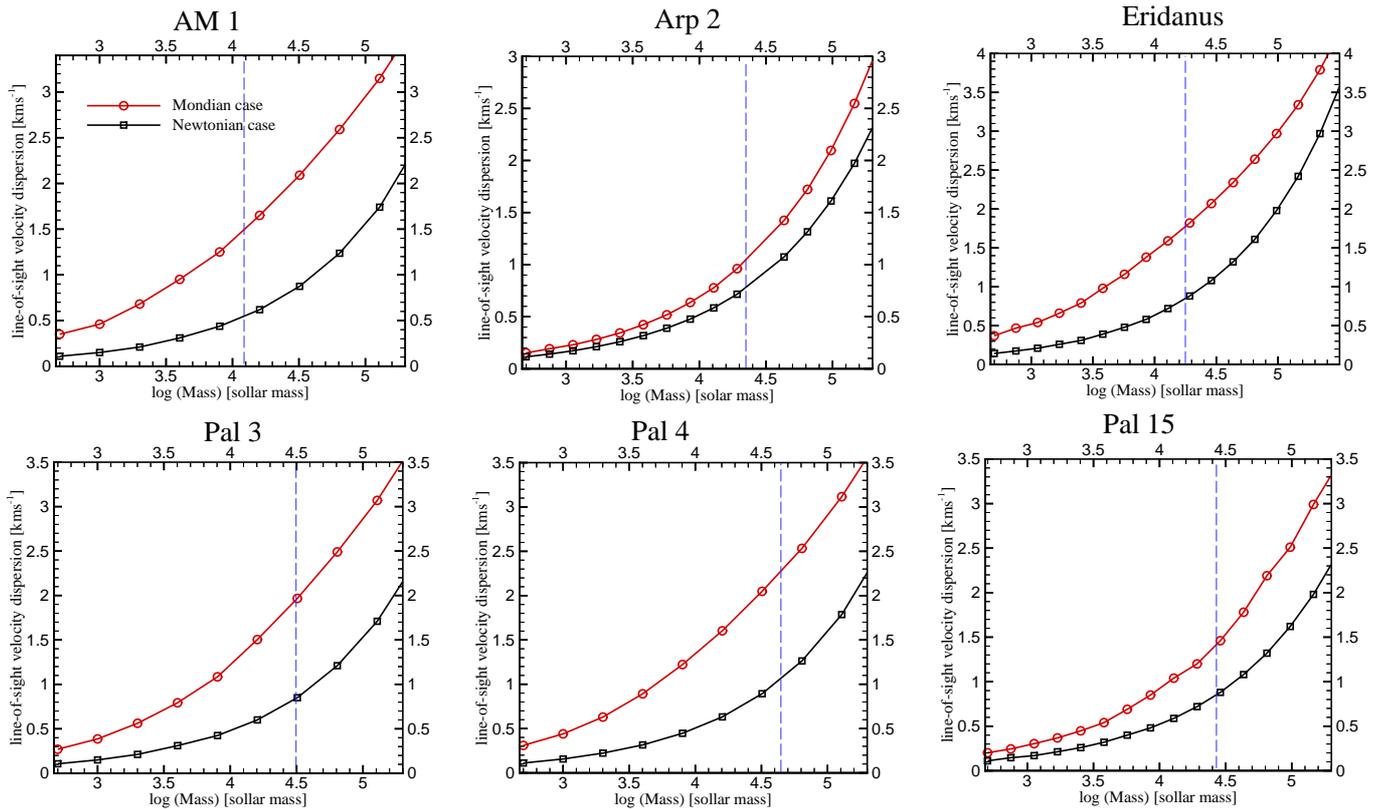}
\caption{Line-of-sight velocity dispersion for clusters listed in
Table 1 for various masses as found by N-MODY. In order to compare
with the real cluster, the half-mass radii of all models are fixed.
For low masses, which mean low internal accelerations, the
prediction of Newtonian dynamics differs from the numerical
solution. As the mass increases (at about $10^6 $M$_{\odot}$), the
internal acceleration grows and the system enters the Newtonian
regime and the numerical solutions get close to the Newtonian
prediction. In order to focus on a reasonable range of mass for
these clusters, we show the results in the range of $500 -
2\times10^{5}$. Dashed vertical lines show the expected cluster
mass, calculated from the absolute luminosity by assuming a stellar
mass-to-light ratio of $M/L_V=2$. This value agrees with observed
mass-to-light ratios of galactic globular clusters (Mieske et al.
2008)} \label{disp}
\end{figure*}

\section{ Modeling the distant globular clusters } \label{S5}

We calculated the velocity dispersion for the Galactic globular
clusters mentioned in Table 1. These clusters are best suited for
testing MOND. A large set of dissipationless N-MODY computations
with MONDian equilibrium initial conditions were performed for
stellar systems that are embedded in the outer MW halo and affected
by different values of the external field. Clusters in Table 1 are
generally far out in the Galactic halo so that the external
acceleration caused by the MW is small. In order to produce
different internal acceleration regimes, we changed the cluster mass
from $500 $M$_{\odot}$ to $10^7 $M$_{\odot}$ and assume the
half-mass radius to be constant. The models are evolved for several
crossing times to reach the equilibrium state. The EFE on the
predicted line-of-sight velocity dispersion for non-isolated stellar
systems with different internal accelerations have been carried out
in HBK09.

For the GCs listed in Table 1, observational efforts are underway to
determine their velocity dispersion and to constrain their mass
(Grebel et al. 2009). Because the half-mass relaxation times of most
clusters in Table 1 are on the order of a Hubble time or even
larger, dynamical evolution does not play an important role for
these clusters, therefore we can assume that the half-light radius
is equal to the half-mass radius. The projected half-mass radii,
$R_{hp}$, and galactocentric distances, $R_G$, are taken from Harris
(1996). The projected half-mass radii relate to the
three-dimensional half-mass radii as $R_{hp}=\gamma R_h$ with
$\gamma \approx 0.74$. For all clusters in Table 1, the tidal radii
are larger than the half-mass radii by a factor of 5 to 10, which
means that tidal effects play no significant role for the internal
dynamics of the clusters in our sample. We perform numerical
modeling to obtain the mean velocity dispersion.

\begin{figure}{}
\begin{center}
\resizebox{9.0cm}{!}{\includegraphics{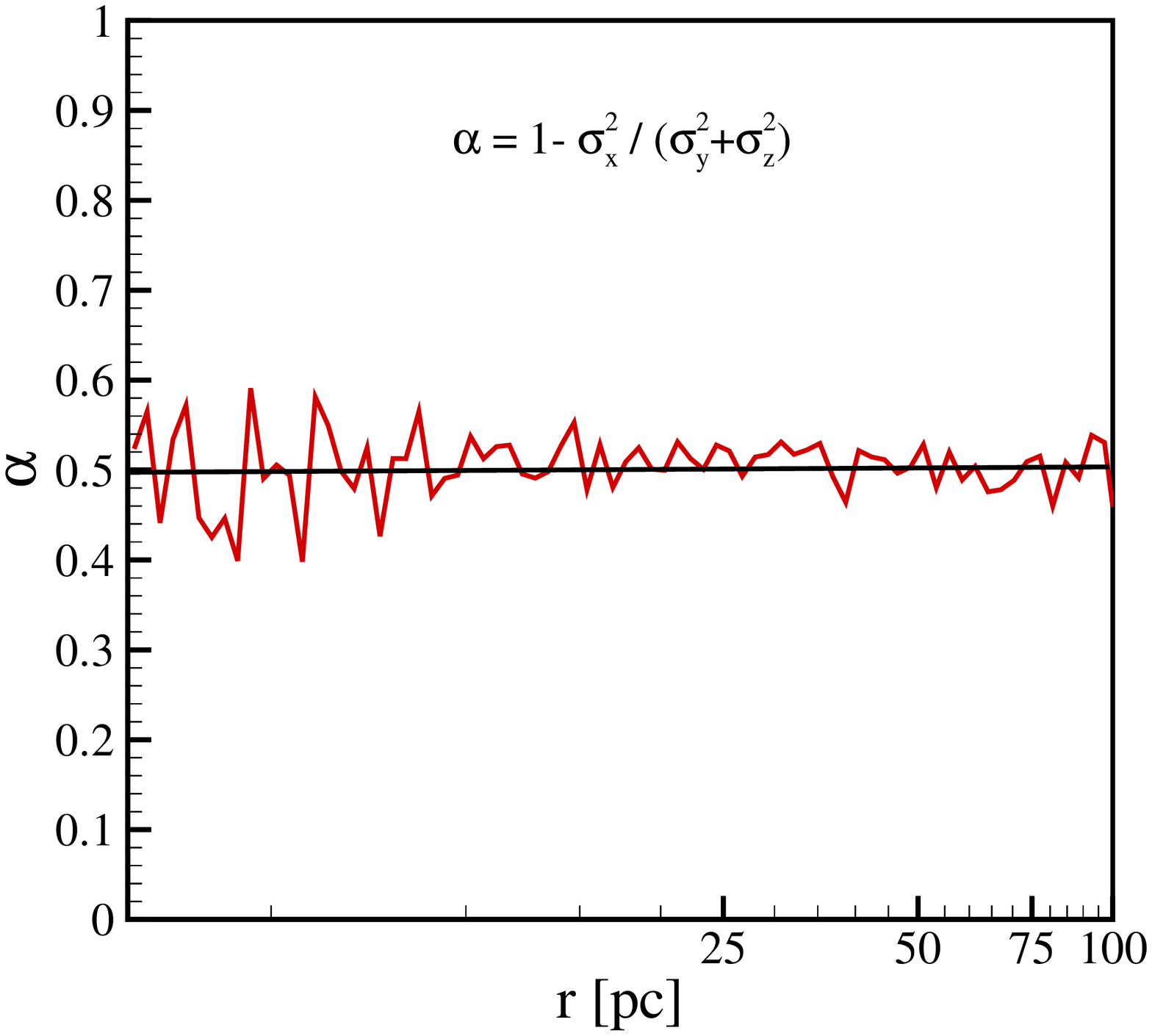}}
\resizebox{9.0cm}{!}{\includegraphics{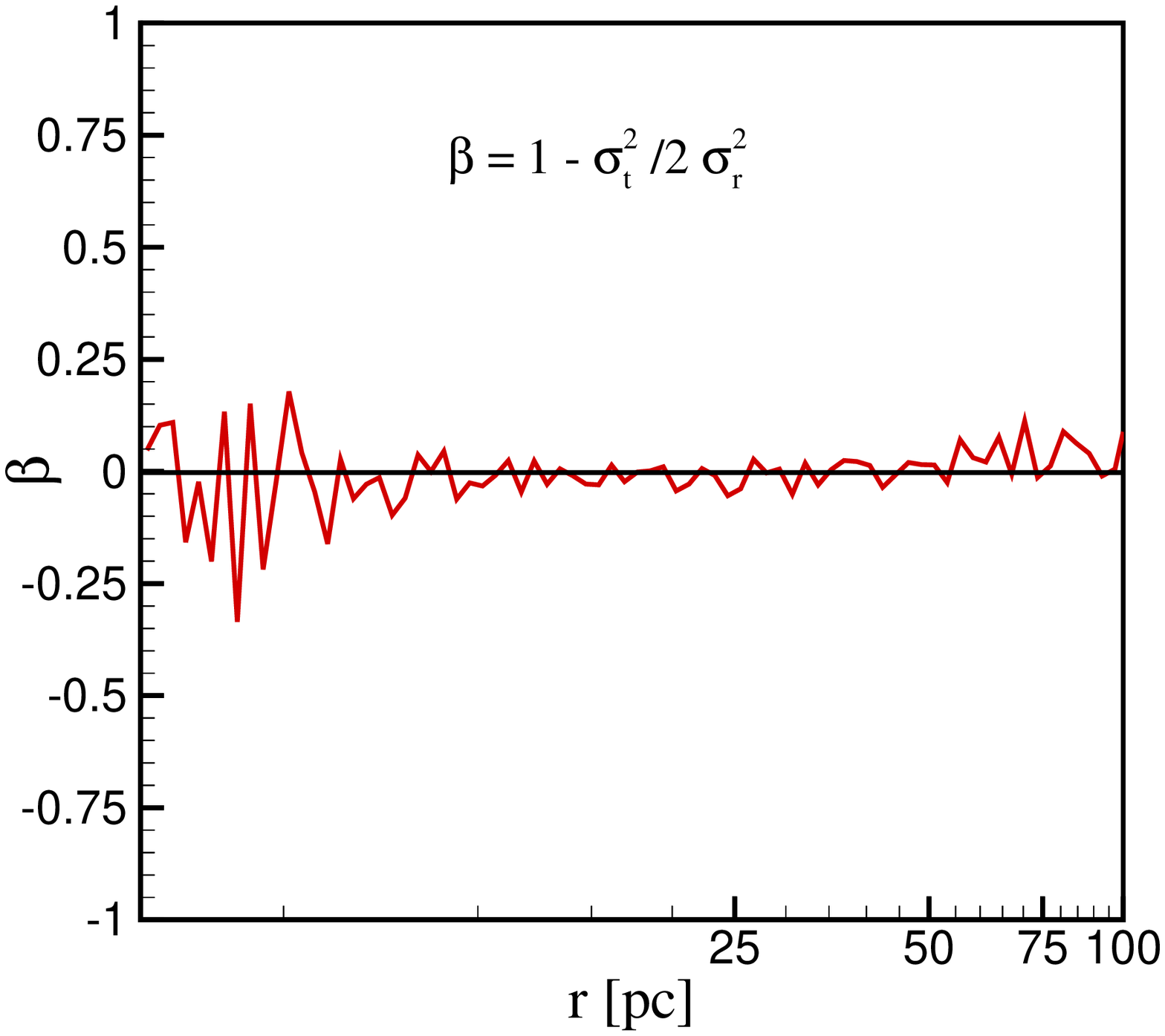}} \caption{
\emph{Top panel:} The ratio of the velocity dispersion in the
direction of the external field compared to the perpendicular
direction for Pal 3 as an example plotted against the radius. The
system is isotropic throughout the cluster. \emph{Bottom panel:} The
anisotropy profile, $\beta$, for Pal 3 as an example plotted against
the radius. The system is isotropic throughout the cluster.
\label{sigmax}}
\end{center}
\end{figure}

\begin{figure}{}
\begin{center}
\resizebox{9.3cm}{!}{\includegraphics{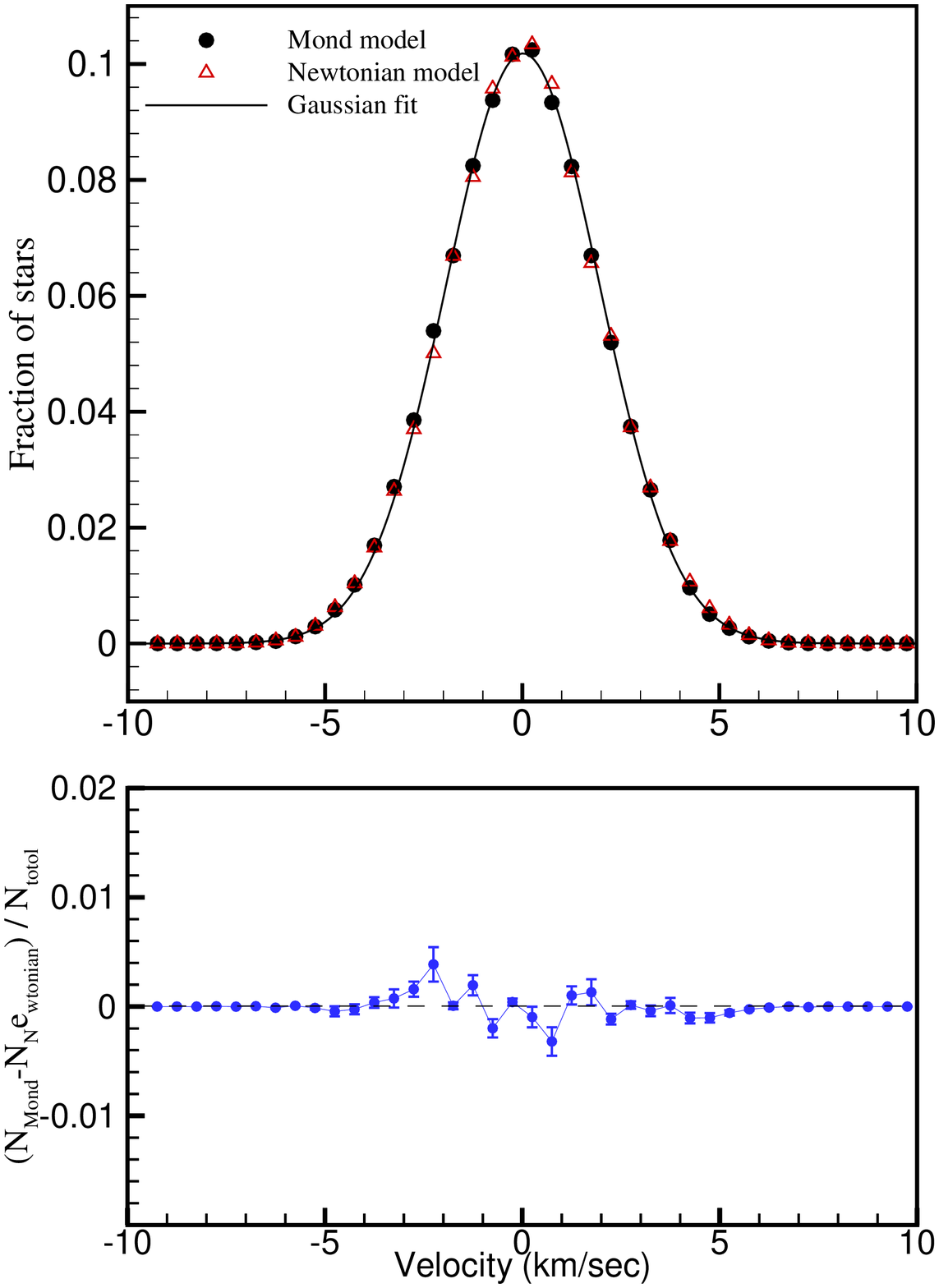}} \caption{
Upper panel: Radial velocity distribution of stars for a Newtonian
(red triangles) and MONDian (black dots) system in which both have
the same internal velocity dispersion. The smoothed line shows a
Gaussian fitted to the distributions. Both distributions follow the
Gaussian fit very closely. Lower panel: The relative difference
between Newtonian and Mondian velocity distributions. There is a
difference between the two models, but it is too small (less than
1\%) to be detected observationally.} \label{gaus}
\end{center}
\end{figure}

Because the cluster masses are not known from observations, it is
useful to calculate line-of-sight velocity dispersions for different
values of cluster mass. If the velocity dispersion is determined
observationally, one can constrain the cluster mass. In Fig. 1 the
resulting global line-of-sight velocity dispersion as a function of
mass is plotted for the six stellar clusters and is compared with
the Newtonian results. The black line shows the Newtonian prediction
for the velocity dispersion, which we calculate using
$\sigma_{LOS,N}=0.36\sqrt{GM/R_h}$ (\cite{hag09}). The red line
shows the numerical calculation of the velocity dispersion in the
MONDian regime. For each cluster the external acceleration of the
Galaxy is given in Table 1. All clusters are in the intermediate
regime, for which there is no analytical prediction in MOND. In the
low-acceleration region, the numerical solutions show a considerable
relative difference with Newtonian results. As the mass increases,
the internal acceleration grows and then gets close to the Newtonian
results (at about $M=10^6$M$_{\odot}$). The vertical dashed lines
show the expected mass for each cluster calculated from the absolute
luminosities by assuming a stellar mass-to-light ratio of $M/L=2$
(\cite{mie08}). It should be noted that in the simulations we do not
assume any $M/L$, but we vary the mass and therefore the $M/L$ ratio
of the clusters. The velocity dispersions corresponding to this mass
for both cases are indicated in Table 1. The $M/L$ ratio of 2 is
only used in Table 1 to illustrate there is a MOND effect, which
makes a difference for the studied clusters. Our predictions for the
MONDian case are lower than those by Baumgardt et al. (2005). The
absolute differences between MONDian and Newtonian results are
largest for the clusters Pal 4, Pal 3, AM 1, and Eridanus. Therefore
these clusters would be the best cases to test MOND. The same
qualitative results were also obtained by Sollima and Nipoti (2009),
who have studied some of the clusters of our list (i.e., Eridanus,
AM1, Pal 3, Pal 4, and Pal 14, which we have studied in HBK09). They
have obtained the cluster's velocity dispersion profile. Their
predictions for the velocity dispersion are generally higher than
ours. For AM1 for example, the mean value of the velocity dispersion
is $\sim$1.7 kms$^{-1}$ since the predicted MOND velocity dispersion
profile is flat throughout the cluster area (Fig. 5 of Sollima and
Nipoti (2009)). This is higher than the value of 1.5 kms$^{-1}$ we
find, which is to be expected because they did not include the EFE.
If we take the EFE into account, the systems tend to be in the
quasi-Newotonian regime, where the gravity is weaker than in the
deep-Mondian regime (HBK09).

In MOND the external field introduces an anisotropy so that clusters
could in principle become elongated. In order to test for a possible
elongation of the velocity dispersion in the direction of the
external field ($\sigma_x$), we define the parameter
$\alpha=1-\sigma_x^2/(\sigma_y^2+\sigma_z^2)$.  Figure \ref{sigmax}
shows that the line-of-sight velocity dispersion in the direction of
the external field differs only randomly compared to that
perpendicular to the external field. This small a difference would
be unobservable.

In order to show the situation of the isotropy in the velocity
dispersion, the anisotropy profile,
$\beta(r)=1-(\sigma_t^2/2\sigma_r^2)$ for one cluster is plotted in
Fig. \ref{sigmax}, where, $ \sigma_t$ is the tangential velocity
dispersion and $\sigma_r$ is the radial velocity dispersion.

Figure \ref{gaus} shows the distribution of radial velocities of the
N$_{total}=10^5$ stars in the velocity range -10 to +10 km/sec for
both Newtonian (red triangles) and Mondian (black dots) cases
together with the best Gaussian fit to this distribution using the
least square method. We created the Newtonian model with a mass of
$177000 $M$_{\odot}$ and the Mondian one with a mass of $32000
$M$_{\odot}$, so that the velocity dispersions are the same. The
bottom panel shows the relative deviation,
(N$_{Newtonian}-$N$_{Mond})/$N$_{total}$, where N is the number of
stars in a velocity bin.  Evidently the observed radial velocity
distribution is well approximated by this Gaussian. The Newtonian
one is slightly different from the Mondian distribution, but the
relative difference is very small (less than 1\%), therefore it is
impossible to find a detectable deviation observationally.

\section{Distinguishing between MOND and Newtonian gravity
models}

Usually, it is not feasible to measure the velocity dispersion of a
large sample of stars in distant and low-mass globular clusters. If
only a few stars are measured, the observed velocity dispersion has
a large error, which makes it difficult to distinguish between MOND
and Newtonian dynamics. Recently, Gentile et al. (2010) have shown
that the small sample size for Pal 14 (i.e., radial velocity of 17
stars were used to determine the observed velocity dispersion) can
only reject MOND with a low confidence level.

\begin{figure*}
\centering
\includegraphics[width=180mm]{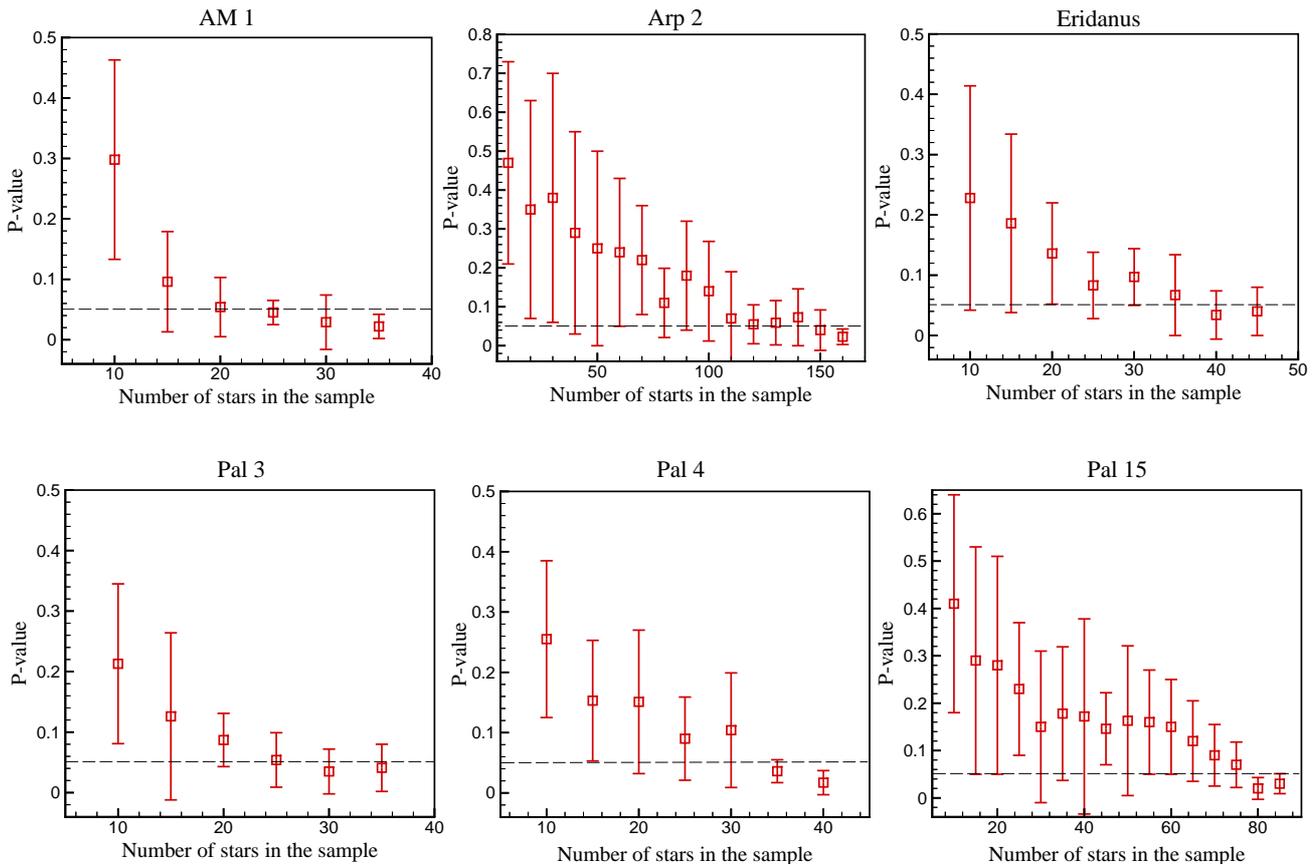}
\caption{ Mean P-value vs. number of stars in the sample for a
number of random realizations of distributions of stars following a
Gaussian with a standard deviation of $\sigma_N$ for clusters listed
in Table 1. The minimum number of stars necessary to reject MOND is
defined where the mean P-value drops below 0.05. }\label{pval}
\end{figure*}

In order to see how many stars are necessary to reject MOND with
high statistical confidence if the underlying velocity dispersion is
Newtonian, we created artificial data sets of increasing sample size
from the Newtonian simulations for each of the six clusters. For
each sample size we produced 20 random realizations of a mock
sample, and each time applied a KS test with the null hypothesis
being that MOND is correct (thus comparing the mock data set from
the Newtonian simulation with the velocity dispersion of the MONDian
simulation, $\sigma_{MOND}$). We approximate the cluster mass from
the V-band luminosity by assuming a stellar mass-to-light ratio of
$M/L=2$ (vertical dashed line in Fig. 1). Figure \ref{pval} shows
the mean P-value from the KS test as a function of the number of
stars in the sample. The number of stars changes in the range of
[10-160]. For any given cluster and a fixed number of stars a
1-sigma range of P-values are shown as error bars. The minimum
number of stars, $N_{min}$, which is necessary to achieve $P<0.05$
(i.e. exclusion confidence of 95\%), are obtained for each cluster
and are given in Table 1. Measurement of $N_{min}$ stars should
therefore be enough to rule out MOND if the cluster mass is
simultaneously determined by independent measurements, possibly
through e.g. star counts.

\section{Conclusions}

We compared global line-of-sight velocity dispersions of six distant
low-density globular clusters of the Milky Way in MONDian and
Newtonian dynamics and showed that they have a significantly higher
velocity dispersion in MOND than the prediction of Newtonian
dynamics.

Using the N-MODY code, we obtained a large set of dissipationless
numerical solutions for globular clusters with the MONDian initial
conditions as end-products of the N-body computations by adding the
external field. In order to produce different internal acceleration
regimes, we changed the mass of the system over a wide range for
each cluster. In addition, our results show that the clusters Pal 4,
Pal 3, AM 1, and Eridanus are the best cases to test MOND owing to
their larger absolute difference between MONDian and Newtonian
velocity dispersions. These results will allow us to test MOND more
rigorously than was possible so far, and this will enable us to
compare MOND with observational data in the future.

Recently Sollima and Nipoti (2009) have performed a test for MOND by
constructing self-consistent dynamical models for outer galactic
clusters. These authors neglected the external field effect of the
Milky Way, and hence obtained higher estimates for the velocity
dispersion of clusters compared with our results. This difference is
reasonable. Indeed, the external field in MOND is leading the system
to move from the deep-MONDian regime toward the quasi-Newtonian
regime, and thus to reduce the velocity dispersion. In other words,
the larger the external field, the smaller the internal
acceleration, which implies a lower velocity dispersion for the
cluster (\cite{hag09}).

It should be noted that the clusters could be on eccentric orbits,
which means that the MOND predictions would be different because of
the variation of the external field (in direction and amplitude)
along the orbit.

Using a KS test, we calculated the minimum number of stars that are
sufficient to exclude MOND (under the hypothesis that these globular
clusters are on circular orbits) at the 95\% confidence level. We
found that between 30 to 80 stars are necessary for most clusters to
distinguish between both cases. This number of stars can be observed
with current $8m$ class telescopes. Additional observational efforts
to determine the velocity dispersions of these clusters and
constraining the mass of the clusters by star counts would be highly
important and provide a strict test of MOND. On the other hand, if
MOND is the correct theory, these observations could be used to
constrain the external field and consequently to put constraints on
the potential in which the systems are embedded. According to the
anisotropy profile, the simulated systems are isotropic throughout.



\begin{acknowledgements}
We would like to thank C. Nipoti for providing us with the N-MODY
code and his help in using it. H.H thanks the stellar dynamics group
of the Argelander Institute for Astronomy for giving financial
support for this research. H.B. acknowledges support from the German
Science foundation through a Heisenberg Fellowship and from the
Australian Research Council through Future Fellowship grant
FT0991052.
\end{acknowledgements}

\end{document}